\documentstyle[aps,psfig,multicol,amsmath,times,prb]{revtex}

\begin{document}

\date{August 3, 2000}
%\draft

\title{Coherent population transfer in coupled semiconductor quantum dots}

\author{Ulrich Hohenester,~$^{*}$
        Filippo Troiani, and Elisa Molinari}
	
\address{Istituto Nazionale per la Fisica della Materia (INFM) and
	 Dipartimento di Fisica, Universit\`a di Modena e Reggio Emilia, 
	 Via Campi 213/A, 41100 Modena, Italy}

\author{Giovanna Panzarini and Chiara Macchiavello}

\address{Istituto Nazionale per la Fisica della Materia (INFM) and
	 Dipartimento di Fisica ``A. Volta'', Universit\`a di Pavia, 
	 Via Bassi 6, 27100 Pavia, Italy}

\maketitle

\begin{abstract} %\baselineskip=2.5ex 

We propose a solid-state implementation of stimulated Raman adiabatic
passage in two coupled semiconductor quantum dots. Proper combination
of two pulsed laser fields allows the coherent carrier transfer between
the two nanostructures without suffering significant losses due to
environment coupling. By use of a general solution scheme for the
carrier states in the double-dot structure, we identify the pertinent
dot and laser parameters.

\end{abstract}


% 78.47.+p  Time-resolved optical spectroscopies and other ultrafast 
%           optical measurements in condensed matter
% 73.40.Gk  Tunneling
% 73.40.Kp  III-V semiconductor-to-semiconductor contacts, p-n junctions, 
%           and heterojunctions
% 78.66.Fd  III-V semiconductors
% 85.30.Vw  Low-dimensional quantum devices (quantum dots, quantum wires
%           etc.)
% 71.35.-y  Excitons and related phenomena
% 73.20.Dx  Electron states in low-dimensional structures (superlattices,
%           quantum well structures and multilayers)

\begin{multicols}{2}
%\narrowtext

Quantum coherence among charge states is known to have strong impact on
the optical properties of confined systems. A striking example of such
coherence is provided by {\em trapped states},~\cite{arimondo:96}\/
where two intense laser fields drive an effective three-level system
into a state which is completely stable against absorption and emission
from the radiation fields. This highly nonlinear coherent-population
trapping is exploited in the process of Stimulated Raman Adiabatic
Passage (STIRAP)~\cite{bergmann:98} to produce complete population
transfer between quantum states: Coherent control of optical
excitations of atoms and molecules, as well as of chemical-reaction
dynamics has been demonstrated in the last
years.~\cite{bergmann:98,gordon:97}

Successful implementations of such schemes in semiconductor
nanostructures would have a strong impetus on the development of novel
opto-electronic devices. However, in contrast to atomic systems carrier
lifetimes in the solid state are much shorter because of the continuous
density-of-states of charge excitations and stronger environment
coupling. Although the possibility of trapped states and coherent
population transfer in semiconductor quantum wells has been
demonstrated theoretically,~\cite{lindberg:95,poetz:97,artoni:00} its
efficiency is expected to be extremely poor as compared to atomic
systems. It has, on the other hand, been noted that the situation could
be much more favorable in semiconductor quantum dots (QDs), where the
strong quantum confinement leads to an atomic-like density of states
and to a strongly suppressed environment coupling.~\cite{poetz:97}
However, in spite of the continuing progress in dot fabrication, all
the available samples still suffer from the effects of inhomogeneity
and dispersion in dot size, which lead to large linewidths when optical
experiments are performed on large QD ensembles. A major advancement in
the field has come from different types of local optical experiments,
that allow the investigation of individual QDs thus avoiding
inhomogeneous broadening,~\cite{single-dot} and simple coherent-carrier
control in single dots has been demonstrated
recently.~\cite{bonadeo:98}

In this Letter, we propose a solid-state implementation of STIRAP in
two coupled semiconductor QDs. Using a general solution scheme for the
carrier states in the QD, which was recently proven successful in
giving a realistic description of experiment,~\cite{hohenester:00a} we
identify the pertinent dot and laser parameters. This will allow us to
argue that an implementation of the proposed scheme should be possible
with the present state-of-the-art sample growth and coherent-carrier
control. In its simplest form, the STIRAP scheme involves three states,
where two long-lived low-energy states $|1\rangle$ and $|3\rangle$ are
dipole coupled to a state $|2\rangle$ of higher energy (which can be
short lived); transitions between states $|1\rangle$ and $|3\rangle$,
on the other hand, are dipole forbidden. Then, by acting with two
appropriately tuned laser pulses on the system, within a STIRAP process
all population is transferred between states $|1\rangle$ and
$|3\rangle$ {\em without ever populating the ``leaky'' state
$|2\rangle$.~\cite{bergmann:98}}\/ This is a remarkable property, as it
allows an almost complete population transfer without suffering
significant losses due to environment coupling of the inter-connecting
level.

Figure~1 sketches the proposed implementation of STIRAP in two coupled
semiconductor QDs: The two dots (with an assumed parabolic in-plane
potential~\cite{dot} and a double-quantum-well confinement along $z$)
are embedded in the intrinsic region of a typical $p$-$i$ field-effect
structure (the dotted line in Fig.~1 indicates the Fermi energy of the
$p$-type region).~\cite{structure} By applying an external gate
voltage, the number of surplus holes in the dot structure can be
precisely controlled; throughout this paper we assume that the double
dot is populated by an additional hole, where further charging is
prohibited because of the Coulomb blockade. If the external electric
field is strong enough, the hole wavefunctions become localized in
either of the two dots ($|R\rangle$ or $|L\rangle$), as can be inferred
from the shaded areas in Fig.~1. In contrast, the electron wavefunction
extends over both dots (see $|e\rangle$ in Fig.~1) because of the
smaller electron mass and the resulting increased inter-dot tunneling
(electron and hole masses are different for most GaAs-based dot
materials,~\cite{hawrylak:98,bimberg:98} e.g., self-assembled
In$_{0.5}$Ga$_{0.5}$As dots). Thus, the inter-connecting state
$|2\rangle$ of the STIRAP process is assigned to the charged-exciton
state (approximately consisting of one electron in state $|e\rangle$
and two holes in states $|R\rangle$ and $|L\rangle$), whereas the
long-lived states $|1\rangle$ and $|3\rangle$ are assigned to the two
hole states.

A detailed theoretical analysis was performed for the dot structure of
Fig.~1, assuming a prototype dot confinement which is parabolic in the
in-plane directions~\cite{dot} and which has a profile along $z$
according to Fig.~1; such confinement potentials have been demonstrated
to be a good approximation for self-assembled quantum dots formed by
strained-layer epitaxy.~\cite{hawrylak:98} In our calculations for the
double dot, we start from the single-particle states derived by solving
the single-particle Schr\"odinger equation within the envelope-function
and effective-mass approximations. For electrons and holes,
respectively, we keep the 20 energetically lowest single-particle
states (i.e., two subbands in the $z$-direction and 10 ``Fock-Darwin''
states of the 2D-harmonic oscillator~\cite{hawrylak:98}) and compute
the charged-exciton states $|X^+\rangle$ through direct diagonalization
of the Hamiltonian matrix (accounting for all possible hole-hole and
electron-hole Coulomb
interactions).~\cite{hohenester:00a,hohenester:00b} Note that within
this scheme the truncation of the Hilbert space is the {\em only}\/
approximation adopted, and that all Coulomb-induced effects are
accounted for in a first-principles manner. For simplicity, in the
following we only consider charged exciton states where the two holes
have parallel spin orientations; experimentally, this can be achieved
by {\em (i)}\/ aligning the spin of the surplus hole by a small
external magnetic field, and {\em (ii)}\/ photo-generating
electron-hole pairs with appropriate spin orientation by use of
circularly polarized light.

\begin{figure}[bp]
\begin{center}
\centerline{\psfig{file=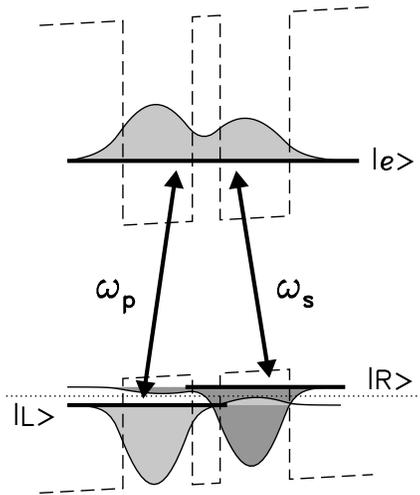,width=2.7in}}
\end{center}
\caption{
Schematic illustration of the proposed double-dot structure. The dashed
lines indicate the confinement potential in the $z$-direction for
electrons (top) and holes (bottom) including an external electric field
of 1.5 mV/nm (the well widths are 5 nm and the distance between the two
dots is 2 nm).$^{\text{\ref{ref:dot}}}$ The hole wavefunction
$|R\rangle$ ($|L\rangle$) is localized in the right (left) dot (shaded
areas; for clarity the energy splitting between the two hole states is
increased here by a factor of 5); because of the lighter electron mass,
the electron wavefunction $|e\rangle$ extends over both dots. The
dotted line indicates the Fermi-energy of the $p$-doped region;
$\omega_s$ ($\omega_p$) indicates the frequency of the Stokes (pump)
laser.}
\end{figure}

Suppose that initially the surplus hole is in state $|h\rangle$ (with
$h=L,R$). The linear absorption spectrum:~\cite{hawrylak:98}

\begin{equation}\label{eq:abs}
  \alpha_h(\omega)\propto\sum_{X^+}|M_{h,X^+}|^2
  {\cal D}_\gamma(\epsilon_h+\omega-\epsilon_{X^+})
\end{equation}

\noindent then provides information about the dipole-allowed
transitions to charged exciton states $|X^+\rangle$; here, $M_{h,X^+}$
is the optical dipole element for the transition between states
$|h\rangle$ and $|X^+\rangle$, $\epsilon_h$ and $\epsilon_{X^+}$ are
the energies of the hole and the charged-exciton states, respectively,
and ${\cal D}_\gamma(\omega)=2\gamma/(\omega^2 +\gamma^2)$ is a
broadened delta function, where the phenomenological damping constant
$\gamma$ accounts for interactions with the dot environment
($\hbar=1$). Figure~2 shows the calculated linear absorption spectra
for the double-dot of Fig.~1, with the surplus hole in the
lowest-energy Fock-Darwin state (with $s$-type symmetry) and:  {\em
(a)}\/ in the right dot; {\em (b)}\/ in the left dot. One observes two
peaks at photon energies of $\sim$130 meV and a peak multiplet at
higher photon energies. A closer inspection of the charged-exciton
states reveals that: {\em (i)}\/ in Fig.~2{\em (a)} (Fig.~2{\em (b)})
the peak at $\omega_p$ ($\omega_s$) corresponds to a transition from
state $|R\rangle$ ($|L\rangle$) to the charged-exciton state of lowest
energy; {\em (ii)}\/ the peaks at higher photon energies correspond to
the transitions to excited charged-exciton states. We thus conclude
that indeed the charged-exciton groundstate optically couples to both
the right-dot and left-dot hole groundstate. Furthermore, since all
excited states are optically well separated from the two peaks at
$\omega_s$ and $\omega_p$, in the following it suffices to consider as
an effective three-level scheme the charged-exciton groundstate only
(henceforth denoted with $|X^+\rangle$), and for the hole state in the
right and left dot, respectively, the Fock-Darwin state of lowest
energy. In this respect two points are worth mentioning: First, our
findings do not depend decisively on the chosen dot parameters, but
rather reflect the basic symmetries of the dot confinement; second,
because of the zero-dimensional nature of carrier states in QDs there
arises no conceptual problems in choosing the Coulomb-renormalized
charged-exciton state as the inter-connecting state required for an
implementation of STIRAP---in contrast to semiconductors of higher
dimensionality, where Coulomb renormalizations can spoil a clear-cut
identification of an effective three-level system.

We next turn to a discussion of the STIRAP process, where the hole is
transferred from state $|R\rangle$ to state $|L\rangle$ by coupling to
the intermediate charged-exciton state $|X^+\rangle$. Suppose that the
system is subjected to two laser pulses, where the frequency of the
``Stokes'' (``pump'') pulse is tuned to the $L$--$X^+$ ($R$--$X^+$)
resonance (see $\omega_s$ and $\omega_p$ in Figs.~1 and 2). To
understand the essence of STIRAP, as a first preliminary step we assume
that the Stokes pulse {\em only}\/ affects the $L$--$X^+$ transition,
whereas the pump pulse {\em only}\/ affects the $R$--$X^+$ transition
(results of our complete time simulations will be presented further
below). Within the interaction picture, the Hamiltonian of the
effective three-level system then reads (neglecting environment
coupling):~\cite{bergmann:98}

\begin{equation}\label{eq:ham}
  {\bf H}=-\frac{\Omega_p(t)}2|X^+\rangle\langle R|-
           \frac{\Omega_s(t)}2|X^+\rangle\langle L|+\mbox{H.c.},
\end{equation}

\noindent with the Rabi frequencies $\Omega_s(t)={\cal
E}_s(t)M_{L,X^+}$, $\Omega_p(t)={\cal E}_p(t)M_{R,X^+}$, and the time
envelopes ${\cal E}_{s,p}(t)$ of the Stokes and pump pulse,
respectively. For time-independent envelopes ${\cal E}_{s,p}$ the
Hamiltonian ${\bf H}$ of Eq.~(\ref{eq:ham}) can be diagonalized:

\begin{equation}
  \begin{array}{lclclcll}
    |a_o\rangle&=&&& \cos\theta|R\rangle&-&\sin\theta|L\rangle,&
    \;\;\;\omega_o=0\\
    |a_\pm\rangle&\propto&|X^+\rangle&\pm(&
    \sin\theta|R\rangle&+&\cos\theta|L\rangle),&\;\;\;
    \omega_\pm=\pm\Omega_{\rm eff},\\
  \end{array}
\end{equation}

\noindent with $|a_o\rangle$, $|a_\pm\rangle$ the eigenvectors and
$\omega_o$, $\omega_\pm$ the eigenvalues, $\theta$ the mixing angle
defined by $\tan\theta=\Omega_p/\Omega_s$, and $\Omega_{\rm eff}=\frac
1 2 (\Omega_s^2+\Omega_p^2)^{\frac 1 2}$. $|a_o\rangle$ has the
remarkable property of being decoupled from the laser fields, i.e., if
the system is initially prepared in this coherent-superposition state
it remains there at any time, since the contribution to the
excited-state amplitude from both hole states interfere destructively.

\begin{figure}
\begin{center}
\centerline{\psfig{file=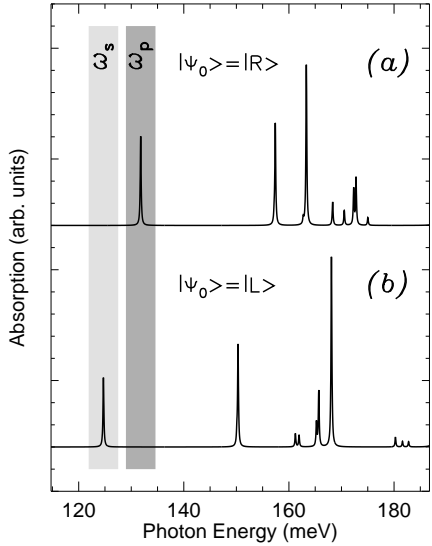,width=2.7in}}
\end{center}
\caption{
Optical absorption spectrum for double-dot structure of Fig.~1 and for
hole in the: {\em (a)}\/ right-dot state $|R\rangle$; {\em (b)}\/
left-dot state $|L\rangle$. Photon energy zero is at the
semiconductor bandgap; $\omega_s$ ($\omega_p$) indicates the photon
energy of the Stokes (pump) laser.
}
\end{figure}

The STIRAP process uses state $|a_o\rangle$ as a vehicle in order to
transfer population between states $|R\rangle$ and $|L\rangle$. Since
$|a_o\rangle$ does not involve contributions from the ``leaky'' state
$|X^+\rangle$, the population can be directly channeled between the two
long-lived hole states (i.e., without suffering losses from
$|X^+\rangle$). Coherent population transfer is achieved by using two
delayed laser pulses which overlap in time (see Fig.~3): Initially, the
system is prepared in state $|R\rangle$. When the first laser pulse
(``Stokes'') is smoothly turned on, the double-dot system is excited at
frequency $\omega_s$. One readily observes from Fig.~2{\em (a)}\/ that
this pulse cannot induce any optical transition. What it does, however,
is {\em (i)}\/ to align the time-dependent state vector
$|\Psi\rangle_t$ with $|a_o\rangle_t=|R\rangle$ (since $\theta=0$ in
the sector $\Omega_s\neq 0$, $\Omega_p=0$), and {\em (ii)}\/ to split
the degeneracy of the eigenvalues $\omega_o$, $\omega_\pm$. Thus, if
the second laser pulse (pump) is smoothly turned on ---such that
throughout $\Omega_{\rm eff}(t)$ remains large enough to avoid
(non-adiabatic) transitions between $|a_o\rangle_t$ and
$|a_\pm\rangle_t$--- all population is transferred between states
$|R\rangle$ and $|L\rangle$ within an adiabatic process where
$|\Psi\rangle_t$ directly follows the time-dependent trapped state
$|a_o\rangle_t$.

\begin{figure}
\begin{center}
\psfig{file=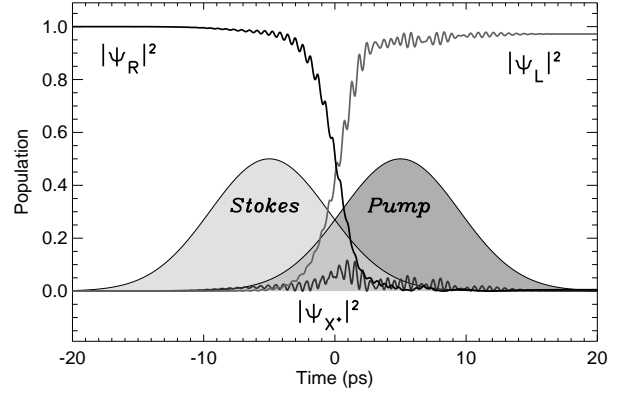,width=3.4in}
\end{center}
\caption{
Time simulation of the STIRAP process which transfers a hole from
the right to the left dot, with $\Psi_h=\langle h|\Psi\rangle$
($h=L,R$) and $\Psi_{X^+}=\langle X^+|\Psi\rangle$. The dashed areas
indicate the time envelopes ${\cal E}_{s,p}(t)$ of the Stokes and pump
pulses (for simulation parameters see Table I).
}
\end{figure}

Figure~3 shows a time simulation of STIRAP accounting for all possible
dipole transitions. The relevant simulation parameters are listed in
Table I; for the time envelopes of the Stokes and pump pulses we use
Gaussians centered at times $\mp 5$ ps, with a full-width of
half-maximum of $\sim$10 ps. In our simulations we use a (relatively
short) effective lifetime of 25 ps for $X^+$ to account for radiative
decay and other decoherence processes due to environment
coupling.~\cite{bonadeo:98} The population transients for the three
states $|R\rangle$, $|L\rangle$, and $|X^+\rangle$ in Fig.~3 show that
for the realistic dot and laser parameters of Table I it is indeed
possible to perform an almost complete population transfer between the
two hole states. We verified that moderate modifications of the dot and
laser parameters do not lead to noticeable changes in the results. Note
that the small population of the leaky state $|X^+\rangle$ in Fig.~3 is
due to the non-adiabatic transients of the laser pulses and to the
coupling to non-resonant transitions; further increase of the Rabi
frequencies was prohibited by the resulting increased losses due to
off-resonant couplings.

In conclusion, we have proposed an implementation of stimulated Raman
adiabatic passage (STIRAP) in a coupled-quantum-dot structure, which is
remote-doped with a single surplus hole. The right-dot and left-dot
hole states are optically connected through the charged-exciton state,
where the spin orientations of electrons and holes are controlled by a
small external magnetic field and by use of circularly polarized light.
Proper combination of the pulsed Stokes and pump laser fields then
allows a coherent population transfer between the two dot states
without suffering significant losses due to environment coupling of the
charged exciton state. Finally, we have discussed that the
implementation of the proposed scheme is experimentally possible with
the available tools of sample growth and coherent-carrier control.

G.~P. thanks Prof. Lucio Claudio Andreani for helpful discussions. This
work was supported in part by the EU under the TMR Network "Ultrafast"
and the IST Project SQID, and by INFM through grant PRA-SSQI.

\begin{table}
\caption{
Parameters used in the time simulation of the STIRAP process shown in
Fig.~3, with: $\epsilon_h$ the single-particle energy of hole state
$|h\rangle$ ($h=L,R$); $\epsilon_{X^+}$ the energy of the
charged-exciton state $|X^+\rangle$ (with respect to the
semiconductor bandgap); $M_{h,X^+}$ the optical dipole elements
for the transition between states $|h\rangle$ and $|X^+\rangle$ (in
units of the dipole matrix element $\mu_o$ of the bulk semiconductor);
$\Omega_s$ ($\Omega_p$) the center-of-pulse Rabi frequency ${\cal
E}_sM_{L,X^+}$ (${\cal E}_pM_{R,X^+}$) of the Stokes (pump) pulse;
$\tau_{X^+}$ the lifetime of state $|X^+\rangle$.
}
  \begin{tabular}{lll}
  $\epsilon_R$		&  $\phantom{1}$25.62		&  meV		\\
  $\epsilon_L$		&  $\phantom{1}$32.71		&  meV		\\
  $\epsilon_{X^+}$	&  157.41			&  meV		\\
  $M_{R,X^+}$		&  $\phantom{11}$0.90		&  $\mu_o$	\\
  $M_{L,X^+}$		&  $\phantom{11}$0.79		&  $\mu_o$	\\
  $\Omega_{s,p}$	&  $\phantom{11}$1.0		&  meV		\\
  $\tau_{X^+}$		&  $\phantom{1}$25.0		&  ps		\\
  \end{tabular}
\end{table}

\end{multicols}

\begin{thebibliography}{10}
\bibitem{dot} The confinement energies due to the in-plane parabolic
potential are $\omega_o^{e}=30$ meV for electrons and $\omega_o^{h}=5$
meV for holes (with this choice, electron and hole wavefunctions have
the same lateral extension); we use material parameters for
GaAs/AlGaAs.  \label{ref:dot}
\end{thebibliography}
\end{document}